\documentclass[preprint,showpacs,preprintnumbers,amsmath,amssymb]{revtex4}
\usepackage{graphicx}% Include figure files
\usepackage{bm}% bold math
\newcommand{\be}[1]{\begin{equation} \label{#1} }
\newcommand{\bea}[1]{\begin{eqnarray} \label{#1} }

\newcommand{\bfi}{\begin{figure}}
\newcommand{\efi}{\end{figure}}
\newcommand{\ee}{\end{equation}}
\newcommand{\eea}{\end{eqnarray}}

\newcommand{\la}{\lesssim}
\newcommand{\ga}{\gtrsim}
\newcommand{\eps}{\epsilon}
\newcommand{\w}{\omega}
\newcommand{\cT}{\mathcal{T}_{_{\rm TE}}}
\newcommand{\gamz}{\gamma^{}_0}
\newcommand{\zetaz}{\zeta^{}_0}
\newcommand{\epsz}{\eps^{}_0}
\newcommand{\muz}{\mu^{}_0}
\newcommand{\Ea}{\mbox{\smash{\shortstack[c]{\vspace{-2.5pt}{\tiny {+}}\\\mbox{$E$}}}}}
\newcommand{\Wa}{\mbox{\smash{\shortstack[c]{\vspace{-2.5pt}{\tiny {+}}\\\mbox{$W$}}}}}
\begin{document}

%\preprint{APS/123-QED}

\title{Plane-wave solutions to frequency-domain and time-domain\\ scattering from magnetodielectric slabs}
\author{Arthur D.Yaghjian and Thorkild B. Hansen}
\address{Consultants, AFRL/SNH, Hanscom AFB, MA 01731, USA}
\date{\today}
\begin{abstract}
Plane-wave representations are used to formulate the exact solutions to frequency-domain and time-domain sources illuminating a magnetodielectric slab with complex permittivity $\epsilon(\omega)$ and permeability $\mu(\omega)$.  In the special case of a line source at $z=0$ a distance $d < L$ in front of an $L$ wide lossless double negative (DNG) slab with $\kappa(\omega^{}_0) =\epsilon(\omega^{}_0)/\epsilon^{}_0 =\mu(\omega^{}_0)/\mu^{}_0 = -1$, the single-frequency ($\omega^{}_0$) solution exhibits not only ``perfectly focused" fields for $z>2L$ but also divergent infinite fields in the region $2d<z<2L$.  In contrast, the solution to the same lossless $\kappa(\omega^{}_0) = -1$ DNG slab illuminated by a sinusoidal wave that begins at some initial time $t =0$ (and thus has a nonzero bandwidth, unlike the single-frequency excitation that begins at $t= -\infty$) is proven to have imperfectly focused fields and convergent finite fields everywhere for all finite time $t$.  The proof hinges on the variation of $\kappa(\omega)$ about $\omega=\omega_0^{}$ having a lower bound imposed by causality and energy conservation. The minimum time found to produce a given resolution is proportional to the estimate obtained by [G\'omez-Santos, Phys. Rev. Lett., {\bf 90}, 077401 (2003)].  Only  as $t \to \infty$ do the fields become perfectly focused in the region $z>2L$ and divergent in the region $2d<z<2L$.  These theoretical results, which are confirmed by numerical examples, imply that divergent fields of the 
single-frequency solution are not caused by an inherent inconsistency in assuming an ideal lossless $\kappa(\omega^{}_0) = -1$ DNG material, but are the result of the continuous single-frequency wave (which contains infinite energy) building up infinite reactive fields during the infinite duration of time from $t=-\infty$ to the present time $t$ that the single-frequency excitation has been applied.  An analogous situation occurs at the resonant frequencies of a lossless cavity.  A single-frequency (zero bandwidth) source inside the cavity produces infinite fields at a resonant frequency, whereas the same source turned on at time $t=0$ (so that it has a nonzero bandwidth) produces finite fields.
\end{abstract}
%                             % Classification Scheme.
%                              %display desired
\maketitle
\section{\label{sec:Introduction}INTRODUCTION}
The main purpose of this paper is to explain and resolve a number of the peculiarities and apparent paradoxes (such as perfectly reproduced source fields as well as divergent fields to the right of the slab) exhibited by the solution to a single-frequency (zero bandwidth) sinusoidal source illuminating a lossless magnetodielectric slab with relative permittivity and permeability equal to negative one \cite{Veselago}--\cite{Wolf}.  This is accomplished by determining the solution for a time-domain (nonzero bandwidth) source produced by turning on the same sinusoidal source  at some finite initial time \cite{Gomez}, \cite{Wolf}, \cite{Z&H}, for example, at $t=0$, and assuming the lossless slab has the slowest possible frequency variation in relative permittivity and permeability (about the value of negative one) allowed by causality and energy conservation.  To clearly reveal the peculiarities and apparent paradoxes in the single-frequency solution, however, we begin by deriving a rigorous plane-wave solution to the 
single-frequency source illuminating a general lossless or  lossy magnetodielectric slab with arbitrary permittivity and permeability.
\section{\label{freq-dom-sol}FREQUENCY-DOMAIN SOLUTION}
The boundary value problem of a time-harmonic ($e^{-i\w t}$, $\w >0$) source illuminating an infinite magnetodielectric slab can be solved simply and rigorously in terms of plane-wave representations \cite{Clemmow}--\cite{H&Y}.  For example, the plane-wave solution for the $x$ component of the electric field $E_x (x,z)$ of  a transverse electric (TE) ($E_y = 0$, $H_y \neq 0$) line source with no variation in the $y$ direction located a distance $d$ in front of a slab (infinite in the $x$ and $y$ directions and normal to $z$) with width $L$, complex permittivity $\eps =\eps' +i \eps''$, and complex permeability $\mu =\mu'+i \mu''$ is given by (see Figure 1)
%
%\mbox{}\\[3mm]\mbox{}
\be{1}
E_x(x,z) =\frac{1}{2\pi}\int\limits_{-\infty}^{+\infty}dh\, e^{ihx} 
\left\{\begin{array}{lll} T_0(h)e^{i\gamma^{}_0 z} +R_0(h) e^{-i\gamma^{}_0 z} &,& 0<z\le d \\
T_s(h)e^{i\gamma z} +R_s(h) e^{-i\gamma z} &,& d\le z \le d+L\\
T(h)e^{i\gamma^{}_0 z}&,& d+L \le z
\end{array}\right.
\ee
where
\begin{subequations}
\label{2}
\be{2a}
\gamma^{}_0 = (k_0^2 -h^2)^\frac{1}{2}\,,\;\;\;\;k_0^2 =\w^2 \mu_0^{}\eps_0^{}
\ee
\be{2b}
\!\!\!\gamma^{} = (k^2 -h^2)^\frac{1}{2}\,,\;\;\;\;k^2 =\w^2 \mu\eps
\ee
\end{subequations}
and $\eps^{}_0$ and $\mu^{}_0$ are the permittivity and permeability of the free space in which the slab is assumed located.  For passive materials $\eps'' \ge 0$ and $\mu'' \ge 0$.  The square root in  the definition (\ref{2a}) of $\gamma_0^{}$ is chosen positive real or positive imaginary depending upon whether $h^2 < k_0^2$ or $h^2 > k_0^2$, respectively.  The sign of the square root in the definition in (\ref{2b}) of $\gamma$ is chosen to keep the imaginary part of $\gamma$ positive.  If $k^2$ is real and $h^2 < k^2$, then $\gamma$ is real and the sign of $\gamma$ is found by inserting a small loss, choosing the imaginary part of $\gamma$ positive, and letting the loss approach zero.  This procedure leads to a positive real $\gamma$ if $\eps$ and $\mu$ are both positive real and a negative real $\gamma$ if $\eps$ and $\mu$ are both negative real. (The signs of the square roots can also be determined from the requirement that the energy flow in the incident and transmitted propagating plane waves and the field decay in the incident and transmitted evanescent plane waves be away from the source.)
\begin{figure}[h]
\mbox{}\\[-.2in]
%\mbox{}\hspace{2.in}
%
%\setbmp{0.25in}{6.in}{4.5in}
%{DNG_Slab_WM_fig1.bmp}
%{c:/pctex/Figures/Figures-Pisa2004/DNG_Slab_WM_fig1.bmp}
%
%\centering
\includegraphics[width =6.in]{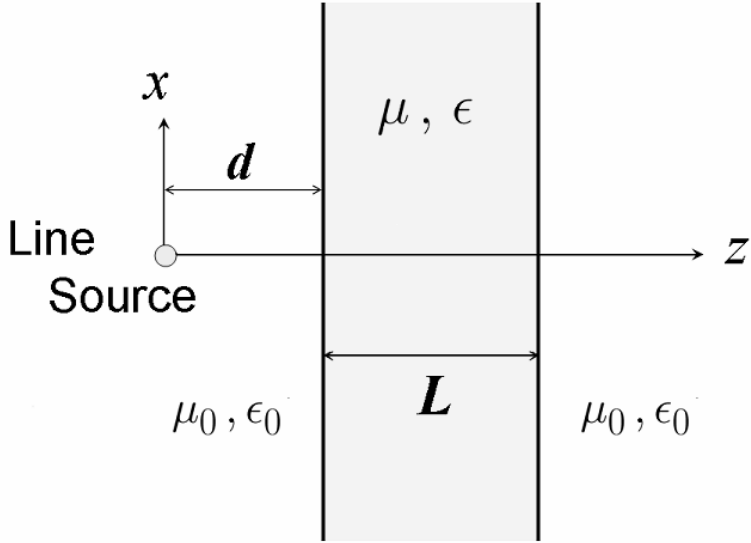}
\mbox{}\\[-2.25cm]
\caption{Geometry of the magnetodielectric slab.}
\end{figure}
%
%\begin{center}\mbox{}\\[-15.mm]
%Figure 1
%\end{center}
\par
The plane-wave spectrum $T_0(h)$ of the fields to the right ($z>0$) of the line source (incident fields) is assumed given.  For example, assume the TE line source is a two-dimensional $y$ directed magnetic line current (magnetization).  Then $T_0(h)$ is independent of $h$ and can be written as
\be{3}
T_0(h) = \frac{E_0}{k_0}
\ee
where $E_0$ is a constant with electric field units and the constant $k_0$ is inserted into the denominator of (\ref{3}) to ensure the dimension of $E_x(x,z)$ is explicitly that of an electric field.
%In general, it can be shown \cite[sec. 3.2.8]{H&Y} that for a source contained within a minimum source region extending to $z=0$
%\be{4}
%\lim_{|h|\to\infty} T_0(h) e^{i\gamma^{}_0 z} \to \infty
%\ee
If at some frequency $\w^{}_0$, the constitutive parameters $\mu(\w^{}_0)/\mu_0^{} =\eps(\w^{}_0)/\eps^{}_0 =-1$, it will be shown that the slab-induced fields of the $y$ directed magnetic-current line source diverge to infinite values in certain regions.  Moreover, this infinite divergence is not peculiar to that particular source.
\par
The reflected spectrum $R_0(h)$ to the left of the slab, the transmitted and reflected spectra, $T_s(h)$ and $R_s(h)$, within the slab, and the transmitted spectrum $T(h)$ to the right of the slab are obtained by equating the tangential components of the electric and magnetic fields across the interfaces of the slab at $z=d$ and $z=d+L$.  Specifically,
\begin{subequations}
\label{5}
\be{5a}
T(h) = T_0(h) \cT(h)
\ee
\be{5b}
T_s(h) =\frac{T(h)}{2} \left( 1+\frac{\eps_0^{}\gamma}{\gamma_0^{}\eps}\right)e^{i(\gamma^{}_0-\gamma)(d+L)} 
\ee
\be{5c}
R_s(h) =\frac{T(h)}{2} \left( 1-\frac{\eps_0^{}\gamma}{\gamma_0^{}\eps}\right)e^{i(\gamma^{}_0+\gamma)(d+L)}
\ee
\be{5d}
R_0(h) = e^{i\gamma_0^{} d}\left( T_s e^{i\gamma d} +R_s e^{-i\gamma d} -T_0 e^{i\gamma_0^{}d} \right)
\ee
\end{subequations}
where the TE transmission coefficient is given as \cite{Pendry}
\be{6}
\cT(h) = \frac{4 e^{-i\gamz L}}{\left(2 +\frac{\eps\gamz}{\epsz\gamma} +\frac{\epsz\gamma}{\eps\gamz}\right)e^{-i\gamma L} +\left(2 -\frac{\eps\gamz}{\epsz\gamma} -\frac{\epsz\gamma}{\eps\gamz}\right)e^{i\gamma L}}\,.
\ee
\boldmath
\subsection{\label{Lossless-slab}Lossless $-1$ double-negative slab}
\unboldmath
In the case of the ``perfectly focusing" slab \cite{Veselago}, \cite{Pendry}, $\eps/\epsz =\mu/\muz =-1$ at some frequency $\w^{}_0$ and we have $\gamma =-\gamz$ if $h^2 < k^2 =k^2_{00}=\w^2_0\muz\epsz$ and $\gamma =\gamz$ if $h^2 >k_{00}^2$.  Then
\be{7}
\cT(h) = e^{-i\gamz 2L}
\ee
and (\ref{5}) become
\begin{subequations}
\label{8}
\be{8a}
T(h) =  T_0(h) e^{-i\gamz 2L}
\ee
\be{8b}
T_s(h) =\left\{ \begin{array}{lllll} T_0(h) e^{i\gamz 2d}&,&\;\; \gamma =-\gamz &,&\;\;h^2 <k^2_{00}\\ 0&,&\;\;\gamma =\gamz &,&\;\;h^2 >k^2_{00} \end{array} \right.
\ee
\be{8c}
R_s(h) =\left\{ \begin{array}{lllll}0&,&\;\;  \gamma =-\gamz &,&\;\;h^2 <k^2_{00} \\ T_0(h) e^{i\gamz 2d}&,&\;\;\gamma =\gamz &,&\;\;h^2 >k^2_{00} \end{array} \right.
\ee
\be{8d}
R_0(h) = 0
\ee
\end{subequations}
so that $E_x(x,z)$ in (\ref{1}) can be written as
\be{9}
E_x(x,z) =\frac{1}{2\pi}\int\limits_{-\infty}^{+\infty}dh\, T_0(h)e^{ihx} 
\left\{\begin{array}{lll} e^{i\gamma^{}_0 z}  &,& 0<z\le d \\
e^{i\gamz (2d- z)}  &,& d\le z \le d+L\\
e^{i\gamz ( z-2L)}&,& d+L \le z\,.
\end{array}\right.
\ee
Since the incident fields are equal to the fields to the right of the source in free space, that is
\be{9'}
E_x^{\rm inc}(x,z) = \frac{1}{2\pi}\int\limits_{-\infty}^{+\infty}T_0(h) e^{i(hx +\gamz z)} dh\,,\;\;\;z>0
\ee
by referring to (\ref{3}), the equations in (\ref{9}) can be re-expressed as
\be{10}
E_x(x,z) =\left\{ \begin{array}{lll} E_x^{\rm inc}(x,z)&,&0<z\le d\\
E_x^{\rm inc}(x, 2d -z)&,& d\le z < 2d\\
\mbox{infinite divergent field}&,& 2d <z<2L\\
E_x^{\rm inc}(x, z-2L)&,& 2L<z
\end{array}\right.
\ee
where from herein out it is assumed that $d<L$.  In other words, the field to the right of the source and to the left of the slab is just the incident field of the source in free space.  The field to the right of the front face of the slab and to the left of $z=2d$ is the image of the source field.  The field between $z=2d$ and $z=2L$ diverges to infinite values.  Most importantly, and quite remarkably, the field in free space to the right of $z=2L$ is just the incident field translated to the right a distance equal to twice the width of the slab.  The phase and magnitude changes of the propagating and evanescent plane waves in the free-space regions between $0<z<d$ and $d+L<z<2L$ are canceled by opposite phase and magnitude changes in the $-1$ double negative (DNG) slab. This perfect replication of the free-space source fields for $z>0$ in the free-space region $z>2L$ to the right of the slab is sometimes referred to as ``perfect focusing" \cite{Veselago}, \cite{Pendry}.
\boldmath
\subsection{\label{Lossy-slab}Lossy $-1$ double-negative slab}
\unboldmath
The solution in (\ref{9})--(\ref{10}) is so unusual that it warrants further investigation.  It is a solution that assumes the loss in the slab (and the surrounding space) is exactly zero.   It may be physically more appealing to insert a small loss into the slab \cite{P&N}, \cite{Smith}, \cite{Shen} and determine the solution for the fields as the loss approaches zero.  This solution can be expressed rigorously from (\ref{1}) in terms of two limits:  the infinite limit for the evanescent spectrum and the limit as the loss in the slab approaches zero.  For simplicity, let the relative loss at the frequency $\w_0^{}$ in $\mu(\w_0^{})$ and $\eps(\w_0^{})$ of the slab be equal and denoted by
\be{10'}
\delta''(\w^{}_0) =\mu''(\w^{}_0)/\muz =\eps''(\w^{}_0)/\epsz >0.  
\ee
Then we can write from (\ref{1})
\be{11}
E_x^{\delta''\to 0}(x,z) =\frac{1}{2\pi}\lim_{\delta''\to 0} \lim_{H\to \infty}\int\limits_{-H}^{+H} dh\, e^{ihx}
\left\{\begin{array}{lll} T_0(h)e^{i\gamma^{}_0 z} +R_0(h) e^{-i\gamma^{}_0 z} &,& 0<z\le d \\
T_s(h)e^{i\gamma z} +R_s(h) e^{-i\gamma z} &,& d\le z \le d+L\\
T(h)e^{i\gamma^{}_0 z}&,& d+L \le z\,.
\end{array}\right.
\ee
One could argue that, in practice, the wavenumber $|h|$ of the evanescent spectrum incident upon the slab should always be truncated to a finite limit $H_0>k_{00}=\w^{}_0\sqrt{\muz\epsz}$ because the evanescent spectrum for $|h|$ greater than some $H_0>k_{00}$ will be lost in the noise.  Then the $\lim_{H\to \infty}\int_{-H}^{+H}$ would be replaced by merely $\int_{-H_0}^{+H_0}$, the $\lim_{\delta''\to 0}$ could be brought under the integral sign, and the solution in (\ref{9})--(\ref{10}) would be approached for $H_0 \gg k_{00}$.
\par
Still, one could ask what the solution becomes if, in principle, the evanescent spectrum is not truncated and the $\lim_{\delta''\to 0}$ is not brought under the integral sign in (\ref{11}).  In that case, we have $\epsilon \gamz/(\epsz \gamma) = -1 +i\delta''(1+k^2_{00}/|\gamz|^2) +O[(\delta'')^2] \approx -1 +i\delta'' + O[(\delta'')^2]$ for the evanescent spectrum if terms in $k_{00}^2/|\gamz|^2$ are neglected compared to unity and we find that $\cT(h)$ in (\ref{6}) for $\delta'' \ll 1$ can be approximated by
\be{11'}
\cT(h) \approx \frac{e^{|\gamz| L}}{\frac{\delta''^2}{4} \, e^{|\gamz|L}+ \, e^{-|\gamz|L}}\,,\;\;\;\;h^2 > k_{00}^2\,.
\ee
This expression reveals that
%$\cT(h)$ is still given approximately by (\ref{7}) for the evanescent part of the spectrum with $|\gamz| < \Gamma_\delta$, but
$\cT(h)e^{-|\gamz|z}$, which comprises the evanescent integrand in (\ref{11}) in the region $z\ge d+L$, rapidly decreases toward zero as $|\gamz|$ grows larger than $\Gamma_\delta$, where $\Gamma_\delta$ is given implicitly by

\begin{subequations}
\label{13}
\be{13aa}
\delta'' \approx e^{-\Gamma_\delta z/2}\,,\;\;\;\; z \ge d+L\,.
\ee
Choosing the minimum value of $z= d+L$, this expression for $\Gamma_\delta$ becomes
\be{13a}
\delta'' \approx e^{-\Gamma_\delta(d+L)/2}
\ee
so that
\be{13b}
\Gamma_\delta \approx -\frac{2}{d+L}\ln \delta''\,.
\ee
\end{subequations}
Moreover, $\cT(h)$ is given approximately by (\ref{7}) for the entire propagating spectrum ($h^2 < k_{00}^2$) with $\delta''\ll 1$, and for the evanescent spectrum up to $\Gamma_\delta$, that is, $0 <|\gamz| <\Gamma_\delta$.  Thus, for the evanescent spectrum in the domain $k_{00}<|h|<H_\delta$, where
\be{13'}
H_\delta \approx \sqrt{\left(\frac{2}{d+L}\ln \delta''\right)^2 +k_{00}^2}
\ee
(\ref{5}) becomes
\begin{subequations}
\label{14}
\be{14a}
T(h) \approx T_0(h)e^{2|\gamz|L}
\ee
\be{14b}
T_s(h) %\approx -i T_0(h) e^{-i\gamz 2L} \delta'' 
\approx -iT_0(h) e^{(2|\gamz| -\Gamma_\delta)L}
\ee
\be{14c}
R_s(h) \approx T_0(h) e^{-2|\gamz| d}
\ee
\be{14d}
R_0(h) \approx -i T_0(h) e^{2|\gamz|(L-d) -\Gamma_\delta L} 
\ee
\end{subequations}
and the evanescent part of the field of (\ref{11}) becomes
\be{15}
E^{{\rm ev},\delta''\to 0}_x(x,z) \approx \lim_{\delta''\to 0}\frac{1}{2\pi} \!\!\!\!\!\!\!\!\int\limits_{k_{00} <|h|<H_\delta}\!\!\!\!\!\!\!\!dh\, T_0(h) e^{ihx} 
\left\{\begin{array}{lll} e^{-|\gamz| z} -i e^{|\gamz| (z+2L-2d) -\Gamma_\delta L} &,& 0<z\le d \\
e^{|\gamz|(z-2d)} -i e^{|\gamz| (2L-z)-\Gamma_\delta L} &,& d\le z \le d+L\\
e^{|\gamz|(2L- z)}&,& d+L \le z
\end{array}\right.
\ee
where $H_\delta$ is given in (\ref{13'}).  For $z\ll \lambda_0$ (where $\lambda_0= 2\pi/k_{00}$ denotes the free-space wavelength),  that is, $z$ extremely close to the source, $H_\delta$ may have to be increased to include all the significant evanescent waves.  However, if $z>2L$, the value of $z$ in (\ref{13aa}) can be chosen equal to its minimum value of $2L$ in that region to obtain 
\begin{subequations}
\be{13a'}
\delta'' \approx e^{-\Gamma_\delta L}\,,\;\;\;\;z>2L
\ee
and $\Gamma_\delta$ in (\ref{13b}) can be replaced by
\be{13b'}
\Gamma_\delta \approx -\frac{1}{L}\ln \delta''\,,\;\;\;\;z>2L
\ee
\end{subequations}
so that $H_\delta$ becomes (see also \cite{P&N}, \cite{Smith}, \cite{Shen})
\be{13''}
H_\delta \approx \sqrt{\left(\frac{1}{L}\ln \delta''\right)^2 +\left(\frac{2\pi}{\lambda_0}\right)^2}\,,\;\;\;\;z>2L\,.
\ee
The propagating spectrum is the same as in (\ref{9}) as $\delta''\to 0$.  Therefore, the propagating spectrum in (\ref{9}) combines with the evanescent spectrum in (\ref{15}) to yield
\be{16}
E_x^{\delta''\to 0}(x,z) =\left\{ \begin{array}{lll} \mbox{bounded field}&,& 0<z< d -(L-d)\\
\mbox{infinite divergent field}&,& d -(L-d)<z<d +(L-d)\\
\mbox{bounded field}&,& d+(L-d)<z<2d\\
\mbox{infinite divergent field}&,& 2d<z< 2L\\
E_x^{\rm inc}(x,z-2L)&,& 2L<z
\end{array}\right.
\ee
where it is assumed in (\ref{16}) that $L/2 < d < L$.  The fields in (\ref{16}) conform to those obtained from the solution in \cite{Milton}, \cite{Milton2}.
\par
Comparing (\ref{16}) and (\ref{10}) reveals that the fields in the region $0<z<2d$ differ depending upon whether the loss $\delta''$ in the slab is made to approach zero before [to get(\ref{10})] or after [to get(\ref{16})] the limit of the integration of the evanescent spectrum is allowed to approach infinity.  However, in the important free-space region to the right of the slab ($z>d+L$), the nature of the fields is independent of the order in which the limit of the loss (approaching zero) and the limit of the integration of the evanescent spectrum (approaching infinity) is taken.  In either case, ``perfect focusing" of the source fields for 
$z>0$ is attained in the limit as the loss approaches zero in the free-space region $z>2L$ to the right of the slab.
\par
Also, in either case, as the loss approaches zero, the field diverges to infinite values in the free-space region $d+L<z<2L$ that lies to the right of the slab.  The fields throughout the free space to the right of the slab ($z>d+L$) may, at first sight, appear to violate the analyticity theorem \cite[ch. V, sec. 4]{C&H}, \cite[Theorem 2.2]{C&K}, which states (to quote \cite[Theorem 2.2]{C&K}), ``If $u$ [our $E_x$ field] is a two times continuously differentiable solution to the [homogeneous] Helmholtz equation in a domain ${\cal D}$ [our region $z>d+L$], then $u$ is analytic." \footnote{Here ``analytic" means with respect to complex $x$ and $z$ in complex neighborhoods containing the real $x$ and $z$ coordinates.}  \mbox{ }This theorem seems to imply that the fields in the free-space region to the right of the slab ($z>d+L$) should be analytic, whereas in fact they are analytic in the region $z>2L$, but diverge to infinite values in the region $d+L<z<2L$ \cite{Maystre}.  This apparent paradox is resolved if it is noted that the analyticity theorem requires that the function be a twice continuously differentiable solution to the homogeneous Helmholtz equation.   It may be possible to weaken this condition to something less restrictive, but certainly the theorem does not apply to fields that diverge to infinite values in part of the region, namely ($d+L<z<2L$) --- as indeed our plane-wave solution demonstrates.\footnote{With reference to (\ref{17}), note that $\lim_{\delta''\to 0} (\nabla^2 +k_{00}^2) E_x^\delta =0$, whereas $(\nabla^2 +k_{00}^2)\lim_{\delta''\to 0} E_x^\delta$ does not exist in the region $d+L<z<2L$ because  $\lim_{\delta''\to 0} E_x^\delta$ does not exist (diverges with infinite oscillation) in that region.} \mbox{ } In particular, an electric field component in a free-space region need not be an analytic function of the spatial coordinates throughout that region if the field is allowed to diverge to infinite values in part of that region.
\par
For a small but nonzero value of the loss parameter $\delta''$, we note from (\ref{15}) [before $\delta''\to 0$ in (\ref{15})] that the fields everywhere to the right of the source ($z>0$) are bounded.  Moreover, with a small but nonzero loss $\delta''$, the fields throughout the free-space region to the right of the slab ($z>d+L$), and throughout the region between the source and the slab ($0<z<d$), are analytic functions of complex $x$ and $z$ in complex neighborhoods of the real $x$ and $z$ coordinates.  For a nonzero loss $\delta''$, equation (\ref{15}) shows that the field in the free-space region $z>d+L$ to the right of the slab is given approximately by
\be{17}
E_x^{\delta}(x,z) \approx \frac{1}{2\pi}\int\limits_ {-H_\delta}  ^{+H_\delta}T_0(h) e^{i[hx +\gamz (z-2L)]} dh\,,\;\;\;z \ge d+L
\ee
with $H_\delta$ given in (\ref{13'}) for $d+L <z < 2L$ and in (\ref{13''}) for $z>2L$.  The transverse resolution just to the right of $z=2L$ can be found by integrating (\ref{17}) for $z=2L$  with $T_0(h)$ for a magnetic-current line source inserted from (\ref{3}) to get
\be{21R1}
E_x^{\delta}(x,2L) \approx \frac{E_0}{\pi k_{00}}\frac{\sin (H_\delta x)}{x}
\ee
which shows that the field of the line source is approximated by a sinc  function with a waist size proportional to $1/H_\delta$.  For two identical line sources separated along the $x$ axis an equal distance $D/2$ from the origin, the $x$ component of the electric field at $z=2L$ is
\be{21R2}
E_x^{\delta ,  D} (x,2L) \approx \frac{E_0}{\pi k_{00}}\left[\frac{\sin [H_\delta(x-D/2)]}{x-D/2} +   \frac{\sin [H_\delta(x+D/2)]}{x+D/2}\right]\,.
\ee
Numerical computations of the field in (\ref{21R2}) of the two sinc functions show that the resolution $\Delta x$, defined as the separation distance such that the two peaks produced by the two sinc functions are 3 dB in intensity above the central minimum, is given by \footnote{The $y$ component of the magnetic field associated with the $x$ component of the electric field in (\ref{21R1}) is given by 
$H_y^\delta (x,2L) =E_0/(2\pi Z_0) \int_{-H_\delta}^{+H_\delta}  e^{ihx}/\gamz  dh$,
which equals $E_0 J_0(k_{00} x)/(2Z_0)$ for $H_\delta =k_{00}$, where $J_0$ is the zeroth order Bessel function and $Z_0 =\sqrt{\muz/\epsz}$ is the impedance of free space.  Numerical computations show that the 3 dB resolution without enhancement ($H_\delta = k_{00}$) determined by this zeroth order Bessel function (rather than by the sinc function) is 
$\Delta x = 1.28 \pi/k_{00} = .64 \lambda_0$.  This value of unenhanced resolution is fairly close to the often stated minimum value of $ .5 \lambda_0$. For $H_\delta \ge k_{00}$, the resolution of $H_y^\delta (x,2L)$ is given approximately by $\Delta x \approx 1.28\pi/H_\delta \approx 1.28\pi/\sqrt{(\ln \delta''/L)^2 +(2\pi/\lambda_0)^2}$, which agrees well with the expression for the minimum $H_y^\delta$ resolution obtained by replacing $\{\epsilon,\mu\}''$ in \cite[eq. (4)]{P&N} with $\{\delta'', \delta''\}$ and solving for $\Delta$ (our $\Delta x$) to get $\Delta x \approx 1.2\pi/\sqrt{[\ln (\delta''/2)/L]^2 +(2\pi/\lambda_0)^2}$.}
\be{18}
\Delta x = \frac{1.53 \pi}{H_\delta} \approx \frac{1.53 \pi}{\sqrt{\left(\frac{1}{L}\ln \delta''\right)^2 +\left(\frac{2\pi}{\lambda_0}\right)^2}}\,. 
\ee
The ``resolution enhancement" $R_e$, defined as the ratio of the resolution with $H_\delta$ to the resolution with the propagating waves alone ($H_\delta = k_{00}$), is thus given simply as
\be{21R4}
R_e =\frac{H_\delta}{k_{00}} \approx \sqrt{\left(\frac{\lambda_0}{2\pi L}\ln \delta''\right)^2 +1}
\ee 
so that 
\be{21R4'}
\Delta x = \frac{1.53 \pi}{k_{00} R_e} = \frac{.76}{R_e} \lambda_0\,.
\ee
For $-\lambda_0 \ln \delta'' /(2\pi L) \gg 1$, (\ref{21R4}) reduces to 
\be{21R4''}
R_e \approx -\frac{\lambda_0}{2\pi L} \ln \delta''
\ee
an expression for resolution enhancement derived by Smith {\em et al.} \cite{Smith}.
The $\delta''$ in (\ref{21R4}) needed to obtain a given resolution enhancement $R_e$ is
\be{19}
\delta'' \approx e^{-2\pi\frac{L}{\lambda_0}\sqrt{R_e^2 -1}}\,.
\ee
For example, to obtain a resolution enhancement of  $R_e = 5$ in a one wavelength slab ($L=\lambda_0$),  the loss parameter $\delta''$ should have a value no larger than about
\be{20}
\delta'' \approx e^{-2\pi\sqrt{24}} \approx 4.3 \times 10^{-14}
\ee
which is an extremely small loss.  For $R_e = 2.5$ resolution enhancement, the loss should be no larger than about
\be{21'}
\delta''  \approx e^{-2\pi\sqrt{(2.5)^2-1}} \approx 5.6 \times 10^{-7}
\ee
quite a small yet more realistic value \cite{Krupka}.  The resolution formulas (\ref{18})--(\ref{21R4'}) are confirmed by the numerical examples in Section \ref{numex}.
\par
As $\delta'' \to 0$ the increasingly large fields in the region $d+L<z<2L$ can be evaluated asymptotically from (\ref{17}), (\ref{13''}), and (\ref{3}) for the $y$ directed magnetic-current line source to get (for $z$ not too close to $2L$)
\be{20'}
E_x^{\delta}(x,z) \stackrel{\delta''\to 0}{\approx} \frac{ E_0}{\pi k_{00} \sqrt{x^2+(2L-z)^2}}
\,\frac{1}{(\delta'')^{2-z/L}} \cos{\left(\frac{x}{L}\ln \delta'' +\tan^{-1}\frac{x}{2L-z}\right)}\,.
\ee
The asymptotic $(\delta'')^{2-z/L}$ decay in (\ref{20'}) agrees with that found for the fields in \cite{Milton}.
\section{\label{Time-dom-sol}TIME-DOMAIN SOLUTION}
In principle, the frequency-domain plane-wave solution to the lossless infinitely long magnetodielectric slab with $\eps(\w^{}_0)/\epsz =\mu(\w^{}_0)/\muz =-1$ has shown that such a slab reproduces the incident fields of a single-frequency (zero bandwidth) source within the free-space region $z>2L$ to the right of the slab.  We have shown that this result holds regardless of whether the loss in the slab material is set equal to zero throughout the formulation of the solution or the loss is chosen nonzero during the formulation of the solution and then allowed to approach zero.  In either case, however, this single-frequency plane-wave solution also predicts that the fields diverge to infinite values in certain regions (defined in (\ref{10}) or (\ref{16})) between $z=d-(L-d)$ and $z=2L$ with $d<L$.
\par
In practice, there will always be some loss in the slab material, some inhomogeneities within the slab material, and some noise level that limits to a finite value the effective wavenumber of the source evanescent waves that reach the front face of the slab. Also, a realistic slab will always be limited to a finite length.  These practical realities will reduce the fields everywhere to finite values.  Nonetheless, it is somewhat disconcerting that the exact classical solution to a source in front of an ideal lossless infinitely long magnetodielectric slab with $\eps(\w^{}_0)/\epsz =\mu(\w^{}_0)/\muz =-1$ has infinite field values in certain regions defined in (\ref{10}) or (\ref{16}).  These infinities may lead one to question the possible existence even within classical physics of an ideal lossless material with $\eps(\w^{}_0)/\epsz =\mu(\w^{}_0)/\muz =-1$ \cite{Garcia}, \cite{Maystre}.
\par
In this section we will illuminate the slab with a time-domain (nonzero bandwidth) sinusoidal wave that turns on at a given finite time in the past (as would any signal in the laboratory) \cite{Gomez}, \cite{Wolf}, \cite{Z&H}, unlike a single-frequency (zero bandwidth) sinusoid that begins in the remote past ($t=-\infty$) and thus illuminates the slab for an infinite amount of time. The former time-domain sinusoid carries a finite amount of energy between the time it turned on and the present time $t$, whereas the latter single-frequency sinusoid carries an infinite amount of energy between the time it turned on ($t=-\infty$) and the present time $t$.  The solution that we obtain in this section to the time-domain sinusoid that is turned on in the finite past reveals that at a finite present time $t$ all fields in the lossless $\eps(\w^{}_0)/\epsz =\mu(\w^{}_0)/\muz =-1$ slab are finite in value, and only as the present time $t\to \infty$ do values of the fields approach infinity in certain regions (the same regions where the single-frequency fields diverge to infinite values).  In other words, it is not the lossless slab with $\eps(\w^{}_0)/\epsz =\mu(\w^{}_0)/\muz =-1$ that inherently leads to the infinitely large single-frequency fields, but the single-frequency continuous wave that illuminates the slab from $t=-\infty$ to the present time $t$ and imparts an infinite amount of reactive energy in certain regions within and near the slab.  The time-domain solution unequivocally explains the origin of the infinite fields encountered in the lossless single-frequency solution.
\par
An analogous situation occurs for a sinusoidal source inside a perfectly conducting (that is, lossless) cavity.  If a single-frequency 
 source is placed inside the cavity, the fields are well-behaved except at the resonant frequencies of the cavity where the fields diverge to infinite values.  One does not conclude from these infinite divergences that a lossless cavity cannot, in principle, exist. One simply acknowledges that the infinite energy in the continuous wave has led to infinite reactive fields in the lossless cavity at the resonant frequencies.  If the single-frequency source is replaced by a time-domain sinusoidal source that begins at a finite time in the past, one finds that the fields inside the cavity remain finite for all finite present time $t$ even at the resonant frequencies.  Only as $t\to\infty$ do the values of the fields of the cavity at the resonant frequencies approach infinity.
\subsection{\label{right-of-slab}Time-domain solution to the right of the slab (\bm{$z\ge d+L$})}
The frequency-domain solution for the TE line source in the region $z\ge d+L$ can be rewritten from (\ref{1}) and (\ref{5a}) as
\be{21}
E_x(x,z) =\frac{1}{2\pi} \int\limits_{-\infty}^{+\infty} T_0(h) \cT(h) e^{i(hx+\gamz z)} dh\,,\;\;\;\;z\ge d+L
\ee
or, alternatively, with the change of integration variable $u=h/\w$
\be{22}
E_x(x,z) =\frac{1}{2\pi} \int\limits_{-\infty}^{+\infty} {\cal T}(u) \cT(\w u) e^{i\w(ux+\zetaz z)} du\,,\;\;\;\;z\ge d+L
\ee
where $\w\zetaz =\gamz(\w u)$ so that
\begin{subequations}
\label{23}
\be{23a}
\zetaz = (1/c^2 -u^2)^\frac{1}{2}\,,\;\;\;\;\muz\epsz=1/c^2
\ee
with $c$ being the speed of light in free space
and
\be{23b}
{\cal T}(u) = \w T_0(\w u).
\ee
\end{subequations}
It is assumed that ${\cal T}(u)$ is independent of the frequency $\w$, for example, as it would be for the $y$ directed magnetic-current line source given in (\ref{3}), that is
\be{24}
{\cal T}(u) =\w T_0(\w u)= \frac{\w E_0}{k_0} = E_0 \,c\,. 
\ee
\par 
Equation (\ref{22}) is the single-frequency ($\w>0$) solution in the region $z\ge d+L$ to the right of the slab for a line source (at $z=0$) with $e^{-i\w t}$ time dependence that has existed from $t=-\infty$ in the remote past.  For a sinusoidal wave, $\cos (\w^{}_0 t)$, that turned on at a finite time $t=-t_0$ in the past and turns off at some future time $t=+t_0$, the frequency-domain spectrum $T_\w$ is given by
\be{25}
T_\w =\frac{1}{2\pi} \int\limits_{-t_0}^{+t_0} \cos (\w^{}_0 t) \;e^{i\w t} dt =\frac{1}{2\pi}\left[\frac{\sin[(\w+\w_0^{})t_0]}{\w+\w^{}_0} +\frac{\sin[(\w-\w_0^{})t_0]}{\w-\w^{}_0}\right]
\ee
in which we assume $t_0 >0$ and $\w_0^{} >0$.  (Choosing the start and end times of the cosine wave equal to $-t_0$ and $+t_0$, respectively, simplifies the time-domain analysis.  At the end of the analysis, the start and end times will be changed to $0$ and $t$, respectively.)
\par
The $x$ component of the time-domain electric field $E_x(x,z,t)$ can now be found by multiplying the integrand in (\ref{22}) by the frequency spectrum $T_\w$ and taking the inverse Fourier transform with respect to $\w$.  Using the fact that $E_x(x,z,t)$ is a real function allows one to integrate over only the positive frequencies in the inverse Fourier transform such that \cite[sec. 5.3]{H&Y}
\be{26}
E_x(x,z,t) =\mbox{Re}\left[\Ea_x(x,z,t) \right]
\ee
where
\be{27}
\Ea_x(x,z,t) =\frac{1}{\pi}\int\limits_0^{+\infty}\int\limits_{-\infty}^{+\infty}T_\w {\cal T}(u)\cT(\w u)e^{i\w(ux+\zetaz z -t)} du\,d\w\,,\;\;\;\;z\ge d+L\,.
\ee
is the analytic-signal time-domain electric field \cite[sec. 5.3]{H&Y}.
Note that as $t_0\to\infty$, the  frequency spectrum $T_\w \to [\delta(\w+\w^{}_0) +\delta(\w-\w^{}_0)]/2$, where $\delta(x)$ denotes the delta function, since
\be{28}
\frac{1}{2\pi}\lim_{t_0\to \infty}\frac{\sin[(\w\pm\w^{}_0)t_0]}{\w\pm\w^{}_0}
= \frac{1}{2} \delta(\w \pm \w^{}_0)\,.
\ee
Thus, for large $t_0$ (\ref{27}) can be expressed approximately as
\be{29}
\Ea_x(x,z,t) \approx \frac{1}{2\pi^2}\int\limits_0^{+\infty}\int\limits_{-\infty}^{+\infty}\frac{\sin[(\w-\w_0^{})t_0]}{\w-\w^{}_0} {\cal T}(u)\cT(\w u)e^{i\w(ux+\zetaz z -t)} du\,d\w\,,\;\;\;\;z\ge d+L\,.
\ee
\par
For large $t_0$, the dominant contribution in (\ref{29}) will come from the $\w$ integration near $\w^{}_0$.  Therefore, the next step toward evaluating the integral in (\ref{29}) analytically is to expand the transmission coefficient $\cT(\w u)$ in a power series about $\w=\w^{}_0$.  Because our primary interest lies in evaluating (\ref{29}) for a lossless slab with relative permittivity and permeability equal to $-1$ at the chosen frequency $\w =\w^{}_0$, that is, $\eps(\w^{}_0)/\epsz =\mu(\w^{}_0)/\muz = -1$, a power series expansion of $\eps(\w)/\epsz$ and $\mu(\w)/\muz$, which are needed in $\cT(\w u)$, can be written as
\begin{subequations}
\label{30}
\be{30a}
\frac{\eps(\w)}{\epsz} = -1 +\frac{1}{\epsz}\frac{d\eps}{d\w}(\w^{}_0) (\w-\w_0^{}) + O\left[(\w-\w^{}_0)^2\right]
\ee
\be{30b}
\frac{\mu(\w)}{\muz} = -1 +\frac{1}{\muz}\frac{d\mu}{d\w}(\w^{}_0) (\w-\w_0^{}) + O\left[(\w-\w^{}_0)^2\right]\,.
\ee
\end{subequations}
For a passive material that is lossless at the frequency $\w_0^{}$ (that is, $\eps''(\w^{}_0) =\mu''(\w^{}_0)=0$), the frequency derivatives $\frac{d\eps''}{d\w}\mbox{\scriptsize$(\w^{}_0)$}$ and $\frac{d\mu''}{d\w}\mbox{\scriptsize$(\w^{}_0)$}$ are also both zero because both $\eps''(\w)$ and $\mu''(\w)$ must be greater than or equal to zero in a passive material, and this is impossible in (\ref{30}) for all $\w$ near $\w^{}_0$ unless $\frac{d\eps''}{d\w}\mbox{\scriptsize$(\w^{}_0)$}$ and $\frac{d\mu''}{d\w}\mbox{\scriptsize$(\w^{}_0)$}$ are both zero.
The coefficients $\frac{d\eps}{d\w}\mbox{\scriptsize$(\w^{}_0)$}/\epsz$ and $\frac{d\mu}{d\w}\mbox{\scriptsize$(\w^{}_0)$}/\muz$ of the $(\w-\w_0^{})$ terms in (\ref{30}) are crucial to the evaluation of the power series for $\cT(\w u)$ and the integral in (\ref{29}). In fact, the final power series expansion for $\cT(\w u)$ obtained below shows that the time-domain solution differs substantially from the single-frequency solution only to the extent that $\frac{d\eps}{d\w}\mbox{\scriptsize$(\w^{}_0)$}/\epsz$ and $\frac{d\mu}{d\w}\mbox{\scriptsize$(\w^{}_0)$}/\muz$ differ from zero. For lossless materials, both causality \cite[sec. 84]{L&L} and energy conservation \cite[app. B]{Y&B} require that these coefficients have the following lower bounds
\be{31}
\frac{1}{\epsz}\frac{d\eps}{d\w}(\w^{}_0) \ge \frac{4}{\w^{}_0}\,,\;\;\;\;\;
\frac{1}{\muz}\frac{d\mu}{d\w}(\w^{}_0) \ge \frac{4}{\w^{}_0}\,.
\ee
Consequently, the $\frac{d\eps}{d\w}\mbox{\scriptsize$(\w^{}_0)$}/\epsz$ and $\frac{d\mu}{d\w}\mbox{\scriptsize$(\w^{}_0)$}/\muz$ that vary the least rapidly near $\w=\w^{}_0$ (and thus will reproduce the source fields most closely in the region $z> 2L$) are given by
\be{32}
\kappa(\w) =\frac{\eps(\w)}{\epsz} =\frac{\mu(\w)}{\muz} = -1 +\frac{4}{\w_0^{}}(\w-\w_0^{}) +O\left[(\w-\w_0^{})^2\right]\,.
\ee
With (\ref{32}) inserted into the expression for $\cT$ in (\ref{6}), one finds that for the propagating waves ($u^2 < \muz\epsz $) it can be approximated by
\be{33}
\cT(\w u) \approx e^{-2i\w\zetaz L}\,,\;\;\;\;u^2 <\muz\epsz =1/c^2\,.
\ee
For the evanescent waves ($u^2 > \muz\epsz $), the quantity $\kappa(\w)\gamz(\w u)/\gamma (\w u) = -1 +4[1+1/(c^2|\zetaz|^2)](\w-\w^{}_0)/\w_0^{} +O\left[(\w-\w_0^{})^2\right] \approx -1 +4(\w-\w^{}_0)/\w_0^{} +O\left[(\w-\w_0^{})^2\right]$ if terms in $1/(c^2|\zetaz|^2)$ are neglected compared to unity and we find
\be{34}
\cT(\w u) \approx \frac{e^{\w|\zetaz| L}}{-\frac{4}{\w^2_0} (\w-\w^{}_0)^2\, e^{\w|\zetaz|L}+e^{-\w|\zetaz|L} }\,,\;\;\;\;u^2 > 1/c^2\,.
\ee
\par
Substitution of $\cT$ from (\ref{33}) and (\ref{34}) into (\ref{29}) gives
\bea{35}
\Ea_x(x,z,t) \approx \frac{1}{2\pi^2}\int\limits_0^{+\infty}d\w\,e^{-i\w t} \left[\,\int\limits_{u^2 < 1/c^2}\!\!\!\frac{\sin[(\w-\w_0^{})t_0]}{\w-\w^{}_0} {\cal T}(u)e^{i\w[ux+\zetaz(z -2L)]}\, du \hspace{2.25cm}\mbox{} \right.\nonumber\\
\left.+\int\limits_{u^2 > 1/c^2}\!\!\!\frac{\sin[(\w-\w_0^{})t_0]\, {\cal T}(u)e^{i\w u x-\w|\zetaz|z}}{(\w-\w^{}_0)\left[e^{-2\w|\zetaz|L}-\frac{4}{\w^2_0} (\w-\w^{}_0)^2 \right]} \, du \right],\;\;z\ge d+L.\hspace{5mm}\mbox{ }
\eea
For large values of $t_0$, the delta-function approximation in (\ref{28}) can be used to evaluate the frequency integration over the propagating spectrum in (\ref{35}).  Specifically, for large $t_0$ and $|t\pm x/c| \ll t_0$ (say $|t\pm x/c|\la t_0/4$), interchanging the order of the $\w$ and $u$ integrations \footnote{The order of integrations in (\ref{35}) can be interchanged because (\ref{35}) is absolutely integrable \cite[p. 37]{J&J}.} reduces (\ref{35}) to
\bea{36}
\Ea_x(x,z,t) \approx \frac{e^{-i\w_0^{} t}}{2\pi}\!\!\! \int\limits_{u^2 < 1/c^2}\!\!\! {\cal T}(u)e^{i\w^{}_0[ux+\zetaz(z -2L)]}\, du \hspace{6.5cm}\mbox{} \nonumber\\
+\frac{1}{2\pi^2}\!\!\!\int\limits_{u^2 > 1/c^2}\!\!\!{\cal T}(u)\int\limits_0^{+\infty}\frac{\sin[(\w-\w_0^{})t_0]\, e^{i\w ux-\w|\zetaz|z}}{(\w-\w^{}_0)\left[e^{-2\w|\zetaz|L}-\frac{4}{\w^2_0} (\w-\w^{}_0)^2 \right]}e^{-i\w t} \, d\w\,du\,,\;\;\;\;z\ge d+L .
\eea
\par
The remaining $\w$ integration in (\ref{36}) can be performed approximately by noting that for $t_0$ large and $|t-ux| \ll t_0$ (say $|t-ux|\la t_0/4$) \footnote{Because the integration of the evanescent spectrum extends to $u=\pm \infty$, it appears that the inequality $|t-ux|\la t_0/4$ cannot be satisfied for all $u$ (unless $x=0$).  However, the following evaluation of the evanescent part of the integrals in (\ref{35}) shows that truncating the $u$ integration to finite limits introduces negligible error.}, its major contribution comes from the integration of $[\sin(\w-\w^{}_0)t_0]/(\w-\w^{}_0)$ between $\w=\w^{}_0 -\pi/t_0$ and $\w=\w^{}_0 +\pi/t_0$, in which domain $[\sin(\w-\w^{}_0)t_0]/(\w-\w^{}_0)$ can be replaced by its average value of approximately $t_0/2$.  Then the $\w$ integration evaluates as
\bea{37}
I &=&\int\limits_0^{+\infty}\frac{\sin[(\w-\w_0^{})t_0]\, e^{i\w ux-\w|\zetaz|z}}{(\w-\w^{}_0)\left[e^{-2\w|\zetaz|L}-\frac{4}{\w^2_0} (\w-\w^{}_0)^2 \right]}e^{-i\w t} \, d\w \nonumber\\
&\approx& \frac{1}{2}t_0\, e^{i\w^{}_0 ux-\w^{}_0|\zetaz| z}\,e^{-i\w_0^{} t}\!\!\!\!\!
\int\limits_{\w^{}_0 -\pi/t_0}^{\w^{}_0 +\pi/t_0}\frac{d\w}{\left[e^{-2\w^{}_0|\zetaz|L}-\frac{4}{\w^2_0} (\w-\w^{}_0)^2 \right]}\nonumber\\
&\approx& -\frac{\w^{}_0 t_0}{4} e^{i\w^{}_0 ux-\w^{}_0|\zetaz| (z-L)} \ln \left|\frac{1-\frac{\w^{}_0 t_0}{2\pi} e^{-\w^{}_0|\zetaz| L}}{1+
\frac{\w^{}_0 t_0}{2\pi} e^{-\w^{}_0|\zetaz| L} }\right|\,. 
\eea
The evanescent spectrum can be truncated at a value $|\zetaz|={\cal Z}_t$ that makes the magnitude of the right-hand side of (\ref{37}) about equal to its value at $\zetaz =0$. To find this value of ${\cal Z}_t$, first approximate the relevant part of the expression in (\ref{37}) by
\be{37approx}
e^{-\w^{}_0{\cal Z}_t (z-L)} \ln \left|\frac{1-\frac{\w^{}_0 t_0}{2\pi} e^{-\w^{}_0{\cal Z}_t L}}{1+\frac{\w^{}_0 t_0}{2\pi} e^{-\w^{}_0{\cal Z}_t L}}\right|
\approx \frac{\w^{}_0 t_0}{\pi} e^{-\w^{}_0{\cal Z}_t z}\,.
\ee
Then we want
\begin{subequations}
\label{38}
\be{38a}
\frac{\w^{}_0 t_0}{\pi} e^{-\w^{}_0{\cal Z}_t z} \approx \ln \left|\frac{1-\frac{\w^{}_0 t_0}{2\pi} }{1+\frac{\w^{}_0 t_0}{2\pi}}\right|
= \ln \left|\frac{1-\frac{2\pi}{\w^{}_0 t_0} }{1+\frac{2\pi}{\w^{}_0 t_0}}\right| \approx \frac{4\pi}{\w^{}_0 t_0}
\ee
which can be solved to give
\be{38aa}
{\cal Z}_t \approx \frac{2}{\w_0^{}z}\ln\left(\frac{\w^{}_0 t_0}{2\pi}\right)\,,\;\;\;\;\w^{}_0 t_0 \gg 2\pi\,.
\ee
Since we are in the region $z\ge d+L$, the value of $z$ can be set equal to its minimum value, $d+L$, in (\ref{38aa}) to obtain
\be{38b}
{\cal Z}_t \approx \frac{2}{\w_0^{}(d+L)}\ln\left(\frac{\w^{}_0 t_0}{2\pi}\right)\,,\;\;\;\;\w^{}_0 t_0 \gg 2\pi\,.
\ee
\end{subequations}
Also, the function $\ln|1- \w^{}_0 t_0\, e^{-\w_0^{}|\zetaz|L}/(2\pi)|$ has such a weak singularity with respect to variation in $u$ as the value of its argument approaches zero that the $u$ integration of (\ref{37}) in (\ref{36}) about this singularity is also negligible compared to the rest of the integration.  Consequently, the $u^2 > 1/c^2$ integration in (\ref{36}) over the evanescent waves can be truncated at $|\zetaz| =\sqrt{u^2 -1/c^2} \approx {\cal Z}_t$ and (\ref{36}) reduces to
\be{39}
\Ea_x(x,z,t) \approx \frac{e^{-i\w^{}_0t}} {2\pi} \int\limits_{|\zetaz| \la \,{\cal Z}_t} {\cal T}(u)e^{i\w^{}_0[ux+\zetaz (z -2L)]} du\,,\;\;|t|\la t_0/4\,,\;\;\;\;\;z\ge d+L\,.
\ee
The condition $|t-ux| \la t_0/4$ has been changed to $|t|\la t_0/4$ in (\ref{39}) because for large $t_0$ the maximum value of $|ux|$ is approximately equal to $|x|{\cal Z}_t$, which is much smaller than $t_0$ if $t_0$ is large enough.
\par
If the time interval that the $\cos (\w^{}_0 t)$ time dependence of the line source is turned on is shifted from $[-t_0, t_0]$ to $[0, 2t_0]$, the condition $|t|\la t_0/4$ in (\ref{39}) shifts to $3t_0/4 \la t \la 5t_0/4$.  Moreover, since (\ref{39}) would then hold for $t\ga 3t_0/4$, causality demands that (\ref{39}) would also hold if the signal turned off at $3t_0/4$ instead of $2t_0$.  In other words, if the line source turned on at $t=0$ with time dependence $\cos (\w^{}_0 t)$, then $t$ can replace $3 t_0/4 \approx t_0$ in (\ref{39}) so that
\be{40}
\Ea_x(x,z,t) \approx \frac{e^{-i\w^{}_0t}} {2\pi} \int\limits_{|\zetaz| \la \,{\cal Z}_t} {\cal T}(u)e^{i\w^{}_0[ux+\zetaz (z -2L)]} du\,,\;\;\;\;z\ge d+L
\ee
with
\be{40Z}
{\cal Z}_t \approx \frac{2}{\w_0^{}(d+L)}\ln\left(\frac{\w^{}_0 t}{2\pi}\right)\,,\;\;\;\;\w^{}_0 t \gg 2\pi\,.
\ee
This equation can be recast in the form of (\ref{17}) by returning the integration variable $u$ to $h/\w_0$ to get
\be{41}
\Ea_x(x,z,t) \approx \frac{e^{-i\w^{}_0t}} {2\pi} \int\limits_{-H_t}^{+H_t} T_0(h)e^{i[hx+\gamz (z -2L)]} dh\,,\;\;\;\;z\ge d+L
\ee
with
\be{42}
H_t = \sqrt{(\w^{}_0{\cal Z}_t)^2 +k_{00}^2}\approx \sqrt{\left[\frac{2}{d+L}\ln\left(\frac{\w_0^{} t}{2\pi}\right)\right]^2 +k_{00}^2}\;.
\ee 
Here, $\gamma^{}_0 = (k_{00}^2 -h^2)^\frac{1}{2}$ and $k_{00}^2 =\w_0^2 \muz\epsz$.  In the region $z>2L$, the value of $z$ can be chosen in (\ref{38aa}) equal to its minimum value of $2L$ in this region, so that $H_t$ in (\ref{42}) is reduced in value to 
\be{42'}
H_t \approx  \sqrt{\left[\frac{1}{L}\ln\left(\frac{\w_0^{} t}{2\pi}\right)\right]^2 +k_{00}^2}\,,\;\;\;\;z>2L\;.
\ee
\par
The field in (\ref{41}) is merely that of the continuous-wave single-frequency field with its evanescent spectrum truncated at $|h| =H_t$ given in (\ref{42}) for $d+L\le z<2L$ and (\ref{42'}) for $z>2L$.  Whereas the evanescent spectrum in (\ref{17}) was truncated because of the presence of a small loss $\delta''$ in the slab, here in (\ref{41}) the field of the lossless slab has its evanescent spectrum truncated because the sinusoidal time dependence of the line source turns on at time $t=0$ instead of $t=-\infty$ and thus has had only a finite amount of time $t$ to generate the evanescent waves.
\par
The resolution $\Delta x$ just to the right of $z=2L$ is now a function of the time $t$ that the source has been turned on.  It is given from (\ref{42'}) as (see also (\ref{18}))
\be{43}
\Delta x \approx \frac{1.53 \pi}{H_t} \approx \frac{1.53 \pi}{\sqrt{\left[\frac{1}{L}\ln\left(f_0 t\right)\right]^2 +\left(\frac{2\pi}{\lambda_0}\right)^2}}
\ee
with a resolution enhancement of
\be{43'}
R_e = \frac{H_t}{k_{00}} \approx \sqrt{\left[\frac{\lambda_0}{2\pi L}\ln\left(f_0 t\right)\right]^2 +1}
\ee
where we have rewritten $k_{00}$ as $2\pi/\lambda_0$ and $\w_0^{}$ as $2\pi f_0$, $\lambda_0$ being the free-space wavelength and $f_0$ the cyclic frequency of the sinusoidal excitation.  To attain a resolution enhancement $R_e$ just to the right of $z=2L$, the line source would have to remain on for a time $t$ given from (\ref{43'}) by
\be{44}
t \approx \frac{1}{f_0} e^{2\pi\frac{L}{\lambda_0}\sqrt{R_e^2 -1}}\,.
\ee
This time is proportional to (and thus critically confirms) the estimate of the time obtained by G\'omez-Santos \cite{Gomez} using a discrete split-frequency approximation ($\pm \Delta \w =\w^{}_\pm -\w^{}_0$) for a narrow band sinusoidal wave.
\par
For example, to attain a resolution of $R_e =5$ at $f_0 = 10$ GHz, a line source illuminating an $L=\lambda_0$ wide slab would have to remain on for a time
\be{45}
t \approx   10^{-10}\; e^{2\pi \sqrt{24}} \approx 39 \mbox{ minutes\,.}
\ee
It is difficult to imagine a one wavelength wide slab made with material having small enough ohmic loss and structural inhomogeneities to maintain the build-up of evanescent fields for 39 minutes at a frequency of 10 GHz.  On the other hand, to attain a resolution of $R_e=2.5$, the same source would have to remain on for
\be{46}
t \approx   10^{-10}\;e^{2\pi \sqrt{(2.5)^2 -1}} \approx 1.8 \times 10^{-4} \mbox{ seconds\,.}
\ee
These results indicate that for a one wavelength slab at a frequency of 10 GHz, a value of resolution enhancement $R_e$  around 2.5 may be feasible just to the right of $z=2L$. The resolution formulas (\ref{43})--(\ref{43'}) are confirmed by the numerical examples in Section \ref{numex}.
\par
Comparing $H_t$ in (\ref{42'}) with $H_\delta$ in (\ref{13''}),  and (\ref{41}) with (\ref{17}), then allows us to determine an asymptotic approximation to (\ref{41}) in the region $d+L<z<2L$ as $t\to \infty$ by replacing $1/\delta''$ in (\ref{20'}) with $\tau =f_0 t$ to obtain for the $y$ directed magnetic-current line source (and for $z$ not too close to $2L$)
\be{46''}
\Ea_x(x,z,t) \stackrel{t\to \infty}{\approx} \frac{E_0\,e^{-i\w^{}_0 t}}{\pi k_{00} \sqrt{x^2+(2L-z)^2}}
\,\tau^{2-z/L} \;\cos{\left(\frac{x}{L}\ln \tau -\tan^{-1}\frac{x}{2L-z}\right)}
\ee
which confirms that the fields diverge to infinite values as $t\to\infty$ in the region $d+L<z<2L$.
\subsection{\label{right-of-source}Time-domain solution throughout the region to the right of the source (\bm{$z > 0$})}
The solution in (\ref{41}) in the region $z\ge d+L$ to the right of the lossless slab shows that the evanescent part of the transmitting spectrum of the slab is truncated by the value $H_t$ given in 
(\ref{42})--(\ref{42'}) that depends upon the amount of time ($t$) that the sinusoidal wave has been turned on (since the initial time $t=0$).  Therefore, for any finite time $t$, the fields of the lossless slab for all $z>0$ to the right of the line source are given approximately by (\ref{9}) with the infinite limits of integration ($-\infty , +\infty$) replaced by ($-H_t, +H_t$).  This means that these fields of the lossless slab are finite everywhere for finite time $t$ and yet approach the fields given in (\ref{10}) as $t\to\infty$.
\par
If a very small loss is inserted into the slab material, an analysis similar to that performed in Section \ref{Lossy-slab} shows that as $t$ gets larger, the fields begin to approach the fields given in (\ref{10}).  As $t$ gets even larger (eventually approaching infinity), however, the fields approach those given in (\ref{16}) with the infinite fields in (\ref{16}) replaced by large finite fields because of the very small but nonzero loss.  After a long enough time $t$, the maximum resolution enhancement for the line source a distance $d<L$ in front of the lossy slab is given just to the right of $z=2L$ by the formula in (\ref{21R4}), that is
\be{47}
R_e \approx \sqrt{\left(\frac{\lambda_0}{2\pi L}\ln \delta''\right)^2 +1}
\ee
where $\delta''$ is the small loss defined as in (\ref{10'}) at the frequency $\w^{}_0$.
\section{Numerical examples}
\label{numex}
\begin{figure}[h]
\mbox{}\\[-2.7cm]
%\mbox{}\hspace{2.in}
%
%\setbmp{0.25in}{6.in}{4.5in}
%{DNG_Slab_WM_fig1.bmp}
%{c:/pctex/Figures/Figures-Pisa2004/DNG_Slab_WM_fig1.bmp}
%
%\centering
\includegraphics[width =6.in]{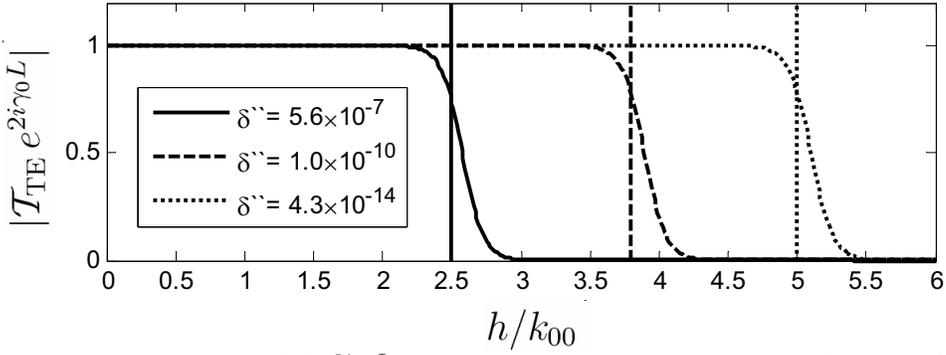}
\mbox{}\\[-1.15cm]
\caption{The magnitude of ${\cal T}_{\rm TE}(h)\,e^{2i\gamma_0 L}$ for a lossy slab as a function of $h/k_{00}$ for $\delta''= 5.6\times 10^{-7}$, $\delta''= 1.0\times 10^{-10}$, and $\delta''= 4.3\times 10^{-14}$. The corresponding values for the resolution enhancements are $R_e= 2.5$, $R_e= 3.8$, and $R_e =5$, as indicated with the vertical lines.}
\label{fig2}
\end{figure}

The formulas for the resolution enhancement will be numerically validated in
this section for both time-harmonic and time-domain line sources. Throughout, the central frequency 
is $f_0=10\mbox{\,GHz}$ and the slab width is $L=\lambda_0$. We consider observation
points only in the region $z\geq 2L$, so the results 
are independent of the distance $d$ between the line source and the slab as long as $0<d<L$.  
We begin with the expression (\ref{21R4}) that determines the resolution enhancement for a 
lossy slab illuminated by a time-harmonic magnetic line source (zero bandwidth). The time-harmonic 
electric field from a single line source at the origin is obtained from (\ref{21}) and (\ref{3}) as
\begin{equation}
\label{num_ex1}
E_x(x,z)=\frac{E_0}{2\pi k_{00}}\int\limits_{-\infty}^{+\infty}{\cal T}_{\rm TE}(h)\,e^{ihx}\,e^{i\gamma_0z}dh,\;\;
z>d+L
\end{equation}
where ${\cal T}_{\rm TE}(h)$ is given in (\ref{6}) with the imaginary components of the
permittivity and permeability satisfying (\ref{10'}). 
Figure \ref{fig2} shows the magnitude of the integrand ${\cal T}_{\rm TE}(h)\,e^{2i\gamma_0L}$ 
(corresponding to $z=2L$) as a function of $h/k_{00}$ for three different values of the loss parameter
$\delta''$  ($k_{00}$ is the free-space propagation constant evaluated at $\omega_0^{}$). 
 The resolution enhancements obtained from (\ref{21R4}) are shown as vertical lines and agree 
well with the observed effective spatial bandwidth of the integrand. 
\par
Next consider a DNG slab that is lossless at the central frequency $f_0=10\mbox{\,GHz}$ and illuminated 
by a magnetic line source with time dependence $v(t)=\sin(\omega_0^{}t)$ for $0\leq t\leq T_e$ and
$v(t)=0$ otherwise. Hence the line source turns on at $t=0$ and turns off at $t=T_e$. We can then use (\ref{num_ex1})
to get the following expression for the analytic-signal time-domain field 
\begin{equation}
\label{num_ex2}
\Ea_x(x,z,t)=\frac{1}{2\pi}\int\limits_{-\infty}^{+\infty} \Wa(h,z,t)\,e^{ihx}\,dh,\;\;
z>d+L
\end{equation}
where 
\begin{equation}
\label{num_ex3}
\Wa(h,z,t)=2E_0\,c\int\limits_{0}^{+\infty}\frac{V_\omega}{\omega} {\cal T}_{\rm TE}(h)\,e^{i\gamma_0z}\,
e^{-i\omega t} \, d\omega
\end{equation}
is the analytic-signal time-domain spectrum and
\begin{equation}
\label{num_ex4}
V_\omega=\frac{1}{4\pi}\left(\frac{e^{iT_e(\omega-\omega_0^{})}-1}{\omega-\omega_0^{}}
-\frac{e^{iT_e(\omega+\omega_0^{})}-1}{\omega+\omega_0^{}}\right)
\end{equation}
is the frequency spectrum corresponding to $v(t)$. (The formula (\ref{num_ex2}) can be obtained directly from
(\ref{27}) with $T_\omega$ in (\ref{25}) replaced by $V_\omega$ in (\ref{num_ex4}).)
We can choose $T_e\,\omega_0^{}=2\pi N$ where $N$ is an integer
to ensure that the ratio $V_\omega/\omega$ is bounded at $\omega=0$. However, in this numerical example
the effective region of integration in (\ref{num_ex3}) is confined to a narrow region centered on $\omega_0^{}$
that does not contain $\omega=0$. The time-domain spectrum (\ref{num_ex3}) is a function of $h$ that
determines the spatial bandwidth of the time-domain field along a line perpendicular to the $z$ axis.
\par
The calculation of the integral  (\ref{num_ex3}) is challenging  because the integrand varies extremely 
rapidly near $\omega_0^{}$. To ensure high accuracy we introduce a frequency-dependent loss that
reduces the width of the region where the values of the integrand are non-negligible. 
This loss manifests itself in the  $(\omega-\omega_0^{})^2$ terms of the expressions for the permittivity and 
permeability (see (\ref{32}))
\begin{equation}
\label{num_ex5}
\frac{\epsilon(\omega)}{\epsilon_0} = \frac{\mu(\omega)}{\mu_0}  = -1 +4 \frac{\omega - \omega_0^{}}{\omega_0^{}}
 + i\left(\frac{1000 (\omega - \omega_0^{})}{\omega_0^{}}\right)^2
\end{equation}
which represents a perfectly lossless $-1$ DNG slab material at $\omega=\omega_0^{}$ that satisfies the 
lower-bound requirements in (\ref{31}). As noted earlier, a passive DNG material that is lossless at 
$\omega=\omega_0^{}$ cannot have losses in the linear term of the power series expansion. 
\par
The following nonuniform $\omega$ discretization is used to compute (\ref{num_ex3}) for observation
points along the line $z=2L+\lambda_0/1000$. We let 
$T_e=10^{-3}\mbox{\,s}$ and evaluate the time-domain spectrum for $0<h/k_{00}<3.5$ using a 
100000-point discretization.  The step length varies as $(\omega-\omega_0^{})^4$ 
for integration points near $\omega_0^{}$. Away from $\omega_0^{}$ the step length is constant. The region 
of integration is $1-10^{-3}<\omega/\omega_0^{} <1+10^{-3}$ for $0<h/k_{00}<2.5$, and  
$1-10^{-9}<\omega/\omega_0^{} <1+10^{-9}$ for $2.5\leq h/k_{00}<3.5$. This discretization
may not be optimal, but it captures the variation of the integrand and ensures that the integral is 
computed accurately. 
\par
\begin{figure}[h]
\mbox{}\\[-1.cm]
%\mbox{}\hspace{2.in}
%
%\setbmp{0.25in}{6.in}{4.5in}
%{DNG_Slab_WM_fig1.bmp}
%{c:/pctex/Figures/Figures-Pisa2004/DNG_Slab_WM_fig1.bmp}
%
%\centering
\includegraphics[width =6.in]{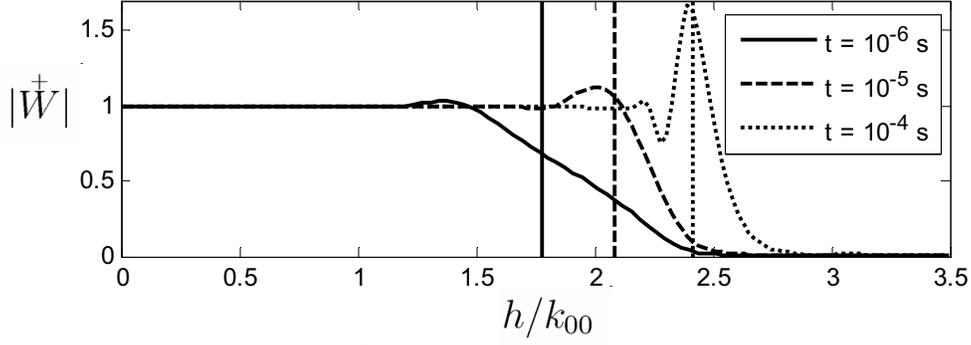}
\mbox{}\\[-1.cm]
\caption{The normalized magnitude of the analytic-signal time-domain spectrum
%$\Wa (h,2L+\lambda_0 /1000,t)$
 as a function of $h/k_{00}$ for $t = 10^{-6}\mbox{\,s}$, $t = 10^{-5}\mbox{\,s}$, and $t = 10^{-4}\mbox{\,s}$. The corresponding values for the resolution enhancements
are $R_e = 1.8$, $R_e = 2.1$, and $R_e =2.5$, as indicated with the vertical lines.}
\label{fig3}
\end{figure}

Figure \ref{fig3} shows the normalized magnitude of the analytic-signal time-domain spectrum $|\Wa (h,2L+\lambda_0 /1000,t)|$ as a function of $h/k_{00}$ evaluated at three different times. Also plotted are the corresponding time-dependent resolution 
enhancements obtained from (\ref{43'}), which are seen to correctly predict the spatial
bandwidth of the time-domain spectrum.
\par
\begin{figure}[h]
\mbox{}\\[-1.cm]
%\mbox{}\hspace{2.in}
%
%\setbmp{0.25in}{6.in}{4.5in}
%{DNG_Slab_WM_fig1.bmp}
%{c:/pctex/Figures/Figures-Pisa2004/DNG_Slab_WM_fig1.bmp}
%
%\centering
\includegraphics[width =6.in]{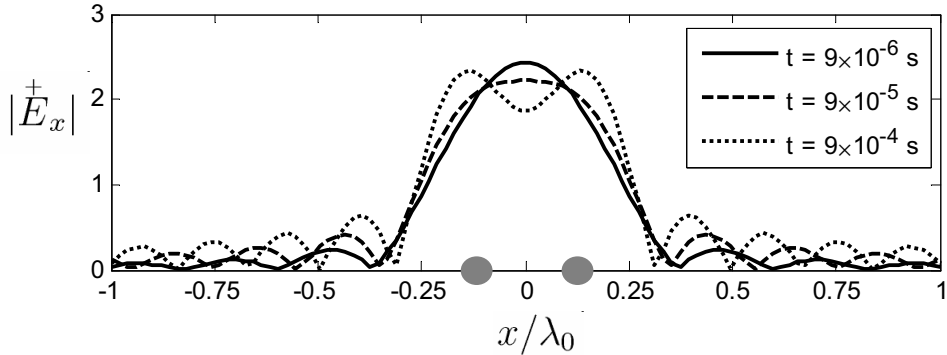}
\mbox{}\\[-1.cm]
\caption{The normalized magnitude of the analytic-signal time-domain electric field %$\Ea_x(x,2L+\lambda_0/1000,t)$ 
from two line sources as a function of $x/\lambda_0$ for $t = 9\times 10^{-6}\mbox{\,s}$, $t = 9\times 10^{-5}\mbox{\,s}$, and $t = 9\times 10^{-4}\mbox{\,s}$, corresponding to a resolution of $\Delta x=0.37\lambda_0$, $\Delta x=  0.32\lambda_0$, and  $\Delta x=   0.28\lambda_0$, respectively. The line sources are located at $(x,z)=(\pm\lambda_0/8,0)$ as indicated by the two gray dots.}
\label{fig4}
\end{figure}

We finally compute the analytic-signal time-domain electric field  from two magnetic line sources that 
are $\lambda_0/4$ apart. Specifically, the line sources are located at  $(x,z)=(\pm\lambda_0/8,0)$. 
The constitutive parameters of the slab are given in (\ref{num_ex5}) and the frequency spectrum 
for the time dependences of the line sources is given by (\ref{num_ex4}). Figure \ref{fig4} shows 
the normalized magnitude of the total analytic-signal electric field $|\Ea_x(x,2L+\lambda_0/1000,t)|$ along the line $z=2L+\lambda_0/1000$ at 
three different times. The positions of the line sources are indicated by two gray dots on the 
$x$ axis of the figure. The resolution $\Delta x$ predicted by (\ref{43}) 
is $\Delta x=0.37\lambda_0$, $\Delta x=  0.32\lambda_0$, and $\Delta x=   0.28\lambda_0$ for 
$t=9\times 10^{-6}\mbox{\,s}$, $t=9\times 10^{-5}\mbox{\,s}$, and 
$t=9\times 10^{-4}\mbox{\,s}$, respectively. Since the actual distance between the line sources is
$\Delta x=0.25\lambda_0$, we should expect  to resolve the line sources only for the later time 
$t=9\times 10^{-4}\mbox{\,s}$. Indeed, the electric field plot shows that the resolution 
improves with time and that the line sources are resolved only for $t=9\times 10^{-4}\mbox{\,s}$. For the two earlier
times the line sources appear as a single peak. Thus, these numerical results confirm that the resolution enhancement
increases with time and that the formula (\ref{43}) gives a good estimate of the spatial resolution as a function of time. 
\newpage 
%
%\mbox{}\\[0cm]
\section{\label{Conclusion}Conclusion}
Plane-wave representations have been used to find the exact solution to a TE line source located in free space a distance $d$ in front of a lossless  or lossy magnetodielectric slab of width $L$ for both single-frequency (zero bandwidth) sinusoidal excitation and for the time-domain (nonzero bandwidth) excitation of a sinusoidal wave that begins at time $t=0$.  In the special case of a lossless slab with relative permittivity and permeability equal to $-1$ ( the $-1$ DNG slab) and $d<L$, the single-frequency source fields incident upon the slab are perfectly reproduced in the free-space region $z>2L$ to the right of the slab, but the fields to the right of the slab in the region $d+L<z<2L$ diverge to infinite values.  These results are shown to hold regardless of whether the loss approaches zero before or after the limits of integration of the evanescent spectrum approach infinity.
\par
In contrast to the single-frequency fields, the time-domain (nonzero bandwidth) fields of the lossless $-1$ DNG slab for the sinusoidal excitation that turns on at $t=0$ (rather than at $t=-\infty$ for the single-frequency excitation) remain finite everywhere for finite present time $t$ and approach the fields of the single-frequency excitation only as $t\to\infty$.  In particular, perfect focusing (zero resolution of the line source) is not attainable after a finite time $t$ because of the nonzero bandwidth \cite{Gomez}, \cite{Wolf}, \cite{Z&H} and the restrictions on the slowest possible variation with frequency of the permittivity and permeability allowed by causality and energy conservation.  These results imply that the divergent infinite fields encountered in the lossless single-frequency solution to the $-1$ DNG slab are caused by the infinite energy in the single-frequency continuous-wave sinusoid that is imparted during the infinite amount of time between $t=-\infty$ and the present time $t$ to the evanescent fields in the vicinity of the slab.  Once the signal is made to turn on at some initial time $t=0$, as would any signal in the laboratory, the fields remain finite everywhere for all future time $t\neq \infty$,  and thus, in principle, there appears to be no inconsistency inherent in postulating an ideal lossless infinitely long magnetodielectric slab with $\eps(\w^{}_0)/\epsz =\mu(\w^{}_0)/\muz = -1$.  In reality, of course, finite losses, inhomogeneities in the material structure, noise levels, and the finite length of the slab, in addition to the nonzero bandwidth of the source coupled with the restrictions imposed by causality and energy conservation, will limit the resolution obtainable in the laboratory.
\par
The major frequency-domain and time-domain theoretical results were  confirmed by the direct numerical computations of Section \ref{numex}.
\begin{acknowledgments}
This work benefitted from discussions with G.W. Milton of the University of Utah and was supported by the U.S. Air Force Office of Scientific Research (AFOSR).
\end{acknowledgments}

\begin{thebibliography}{99}
%
\bibitem{Veselago}  V.G. Veselago,  Sov. Phys. Usp.  {\bf 10}, 509 (1968).
%
\bibitem{Pendry}  J.B. Pendry, Phys. Rev. Lett. {\bf 85}, 3966 (2000).
%
\bibitem{Garcia} N. Garcia and M. Nieto-Vesperinas,  Phys. Rev. Lett. {\bf 88}, 207403 (2002); Erratum, {\bf 90}, 229903 (2003).
%
\bibitem{Gomez}  G. G\'omez-Santos, Phys. Rev. Lett. {\bf 90}, 077401 (2003).
%
\bibitem{Maystre} D. Maystre and S. Enoch, J. Opt. Soc. Am. A {\bf 21}, 122 (2004).
%
\bibitem{Wolf} D.A. de Wolf, IEEE Trans. Antennas Propagat. {\bf 53}, 270 (2005); {\bf 54}, 263 (2006).
%
\bibitem{Z&H}  R.W. Ziolkowski and E. Heyman, Phys. Rev. E {\bf 64}, 056625 (2001).
%
\bibitem{Clemmow} P.C. Clemmow, {\em The Plane Wave Spectrum Representation of Electromagnetic Fields} (Pergamon, Oxford UK, 1966).
%
\bibitem{Kerns} D.M. Kerns,  {\em Plane-Wave Scattering-Matrix Theory of Antennas and Antenna-Antenna Interactions} (U.S. Government Printing Office, Washington DC,  1981).
%
\bibitem{Kong} J.A. Kong,  {\em Electromagnetic Wave Theory} (Wiley, New York, 1986).
%
\bibitem{Kong2} J.A. Kong, PIER {\bf 35}, 1 (2001).
%
\bibitem{H&Y} T.B. Hansen and A.D. Yaghjian,  {\em Plane-Wave Theory of Time-Domain Fields: Near-Field Scanning Applications} (IEEE/Wiley, Piscataway NJ,  1999).
%
\bibitem{Milton} G.W. Milton {\em et al.}, Proc. Roy. Soc. A  {\bf 461}, 3999 (2005).
%
\bibitem{Milton2} V.A. Podolskiy, N.A. Kuhta, and G.W. Milton, Appl. Phys. Lett. {\bf 87}, 231113 (2005).
%
\bibitem{C&H} R. Courant and D. Hilbert, {\em Methods of Mathematical Physics} (Interscience, New York, 1953).
%
\bibitem{C&K} D. Colton and R. Kress, {\em Inverse Acoustic and Electromagnetic Scattering Theory} (Springer-Verlag, New York, 1992).
%
\bibitem{P&N} V.A. Podolskiy and E.E. Narimanov, Optics Letters {\bf 30}, 75 (2005).
%
\bibitem{Smith} D.R. Smith {\em et al.},  Appl. Phys. Lett. {\bf 82}, 1506 (2003).
%
\bibitem{Krupka} J. Krupka {\em et al.},  Meas. Sci. Technol. {\bf 10}, 387 (1999).
%
\bibitem{L&L} L.D. Landau, E.M. Lifshitz and L.P. Pitaevskii, {\em Electrodynamics of Continuous Media, 2nd ed.}  (Butterworth-Heinemann, Oxford UK,  1984).
%
\bibitem{Y&B} A.D. Yaghjian and S.R. Best,  IEEE Trans. Antennas Propagat. {\bf 53}, 1298 (2005).
%
\bibitem{J&J} H. Jeffreys and B.S. Jeffreys, {\em Methods of Mathematical Physics, 3rd ed.} (Cambridge University Press, Cambridge UK, 1956).
%
\bibitem{Shen} L. Shen and S. He,  Phys. Letts. A {\bf 309}, 298 (2003).
%
\end{thebibliography}
\end{document}